\documentclass[aps,pra,superscriptaddress,showpacs]{revtex4}
\usepackage{amssymb,bm}
\usepackage{graphicx}
\usepackage{amsmath}

\begin{document}

\title{Energy dependence of $e^+e^- \to 6\pi$
and $e^+e^- \to N\bar{N}$ cross sections near the $N\bar{N}$ threshold}

\author{A.E. Obrazovsky}\email{A.E.Obrazovsky@inp.nsk.su}
\affiliation{Budker Institute of Nuclear Physics SB RAS, 630090 Novosibirsk, Russia}
\author{S.I. Serednyakov}\email{S.I.Serednyakov@inp.nsk.su}
\affiliation{Budker Institute of Nuclear Physics SB RAS, 630090 Novosibirsk, Russia} 
\affiliation{Novosibirsk State University, 630090 Novosibirsk, Russia}

\date{\today}

\begin{abstract}
Using recent BABAR, CMD-3 and SND data, the sum of $e^+e^- \to 3(\pi^+\pi^-), 2(\pi^+\pi^-\pi^0), p\bar{p}, n\bar{n}$
cross sections is obtained. Unlike $e^+e^- \to 3(\pi^+\pi^-)$ and $e^+e^- \to 2(\pi^+\pi^-\pi^0)$ processes,
no structures in total cross section are found near the $N\bar{N}$ threshold within the limits of measurement errors.
\end{abstract}


\maketitle

First observation of six-pion production in electron-positron annihilation
was performed in $\mu\pi$ experiment at ADONE collider~\cite{mp}.
The dip in $e^+e^- \to 3(\pi^+\pi^-)$ cross section near the $N\bar{N}$ threshold
was found in DM1 experiment at DCI~\cite{dm1}. Later, it was confirmed
by more precise measurements in FOCUS~\cite{foc} and BABAR~\cite{b6pi} experiments.
The similar structure also was observed in $e^+e^- \to 2(\pi^+\pi^-\pi^0)$ mode~\cite{b6pi, bal}.
One of the possible theoretical explanations of this dip 
suggests the existence of $p\bar{p}$ underthreshold bound state~\cite{nntheory}.

Recently, more precise BABAR and SND data on $e^+e^- \to p\bar{p}$~\cite{bpp} 
and $e^+e^- \to n\bar{n}$~\cite{snn} cross sections were published together with the
new CMD-3 data on $e^+e^- \to 3(\pi^+\pi^-)$~\cite{c6pic} and
$e^+e^- \to 2(\pi^+\pi^-\pi^0)$ cross sections~\cite{c6pin}.

We suggest that the sum of $e^+e^-\to hadrons$ cross sections including $N\bar{N}$ final state
should contain no structure near the $N\bar{N}$ threshold due to absence of quark-antiquark resonances
in this energy range.

For the purposes of this work only $e^+e^- \to 3(\pi^+\pi^-)$, $e^+e^- \to 2(\pi^+\pi^-\pi^0)$ 
and $e^+e^- \to N\bar{N}$ processes can be considered,  as far as the cross sections of
$e^+e^- \to 2\pi, KK, 3\pi, KK\pi, 4\pi, KK\pi\pi, 5\pi, {\ldots}$ don't contain structures
near the $N\bar{N}$ threshold~\cite{bhad}. We use CMD-3 data on 
$e^+e^- \to 6\pi$ processes, BABAR and SND data on $e^+e^- \to N\bar{N}$ processes.
The sum of isovector $e^+e^- \to 3(\pi^+\pi^-)$ and $e^+e^- \to 2(\pi^+\pi^-\pi^0)$
cross sections with isoscalar $e^+e^- \to \eta\pi^+\pi^-\pi^0$ background subtraction~\cite{e3p} is shown in fig.1a.
The sum of $e^+e^- \to p\bar{p}$ and $e^+e^- \to n\bar{n}$ cross sections is shown in fig.1b.
These sums are fitted by constants in the energy ranges $2E$ = 1.85--1.876 GeV and 1.876--1.92 GeV.
The cross section discontinuity values are: $\Delta\sigma$ = -1.5 $\pm$ 0.2 nb 
for the sum of $e^+e^- \to 6\pi$ processes, $\Delta\sigma$ = 1.6 $\pm$ 0.25 nb
for the sum of $e^+e^- \to N\bar{N}$ processes. For the total sum of $e^+e^- \to 6\pi$
and $e^+e^- \to 6\pi$ processes $\Delta\sigma$ = 0.1 $\pm$ 0.3 nb (fig. 1c).
The quoted uncertainties include
both statistical and systematic errors. The total cross section
doesn't contain structures within measurement error limits.

We suggest that negative discontinuity in the cross section of $e^+e^- \to 6\pi$ processes
at the $N\bar{N}$ threshold can be quantitatively explained by the opening of new annihilation
channel $e^+e^- \to N\bar{N}$. However, it is unclear, why it is isovector $e^+e^- \to 6\pi$ final state  
that provide nearly full compensation in the total cross section $e^+e^- \to hadrons$.

The authors are grateful to V.\,P.\,Druzhinin and A.\,I.\,Milstein
for useful discussions. The work is partially supported by
the Ministry of Education and Science of the Russian
Federation, scientific school grant NSh-2479.2014.2 and 
Russian Fund for Basuc Research grant RFBR 12-02-00065-a.

%
\begin{figure}
\includegraphics{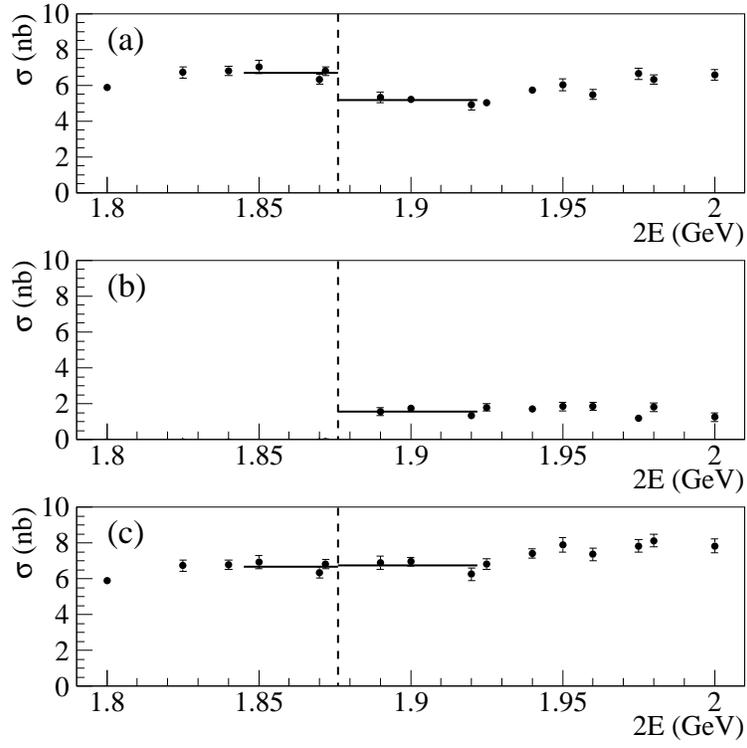}
\caption{(a) The sum of $e^+e^- \to 3(\pi^+\pi^-)$
and $e^+e^- \to 2(\pi^+\pi^-\pi^0)$ cross sections 
with $e^+e^- \to \eta\pi^+\pi^-\pi^0$ background subtraction; (b)
The sum of $e^+e^- \to p\bar{p}$ and $e^+e^- \to n\bar{n}$ cross sections;
(c) The (a)+(b) sum. Only statistical errors are shown. The dashed line is the $N\bar{N}$ threshold.}
\end{figure}

\end{document}